\documentclass[conference]{IEEEtran}

\usepackage{cite}      
\usepackage{graphicx}  
\usepackage{psfrag}    
\usepackage{subfigure} 
\usepackage{url}       
\usepackage{stfloats}  
\usepackage{amsmath}   
\usepackage{amssymb}   

\usepackage{amsmath}
\usepackage{amsthm}
\usepackage{amssymb}
\usepackage{cite}
\usepackage{graphicx}
\usepackage{booktabs}
\usepackage{subfigure}

\newcommand{\hf}{\frac{1}{2}}

\newcommand{\q}{\mathsf{q}}

\newtheorem{theo}{Theorem}

\begin{document}
%
\title{On Noisy Network Coding for a Gaussian Relay Chain Network with
Correlated Noises}

\author{Lei Zhou and Wei Yu \\
       Department of Electrical and Computer Engineering, \\
       University of Toronto, Toronto, Ontario M5S 3G4, Canada \\
       emails: \{zhoulei, weiyu\}@comm.utoronto.ca
}

\maketitle

\begin{abstract}
Noisy network coding, which elegantly combines the conventional
compress-and-forward relaying strategy and ideas from network coding,
has recently drawn much attention for its simplicity and optimality in
achieving to within constant gap of the capacity of the multisource
multicast Gaussian network. The constant-gap result, however, applies
only to Gaussian relay networks with independent noises.  This paper
investigates the application of noisy network coding to networks with
correlated noises. By focusing on a four-node Gaussian relay chain
network with a particular noise correlation structure, it is shown
that noisy network coding can no longer achieve to within constant gap
to capacity with the choice of Gaussian inputs and Gaussian
quantization. The cut-set bound of the relay chain network in this
particular case, however, can be achieved to within half a bit by a
simple concatenation of a correlation-aware noisy network coding
strategy and a decode-and-forward scheme.
\end{abstract}

\section{Introduction}

The capacity region of the Gaussian relay network has been open for decades.
Recently, the capacities of several relay networks with simple structures have
been approximated to within constant number of bits.
For example, for the three-node Gaussian relay channel, Avestimehr and
Tse \cite{Avestimehr_relaynetwork} showed that the decode-and-forward
strategy achieves to within half a  bit of the capacity; Chang, Chung,
and Lee \cite{Chang_constantbitrelay} proved that the
compress-and-forward rate is within half a bit of the capacity, and the
amplify-and-forward rate is within one bit.

In their breakthrough work, Avestimehr and Tse
\cite{Avestimehr_relaynetwork} further showed that, the capacity of
the single-source single-destination Gaussian relay network in general
can be achieved to within constant bits via a universal relaying
scheme called quantize-map-and-forward (QMF). They also showed that,
the gap to capacity is only related to the number of nodes in the
network.

Parallel to Avestimehr and Tse's work, Lim, Kim, El Gamal, and Chung
\cite{Kim_noisy_network_coding} proposed a noisy network coding
strategy that naturally extends the conventional compress-and-forward
scheme of Cover and El Gamal \cite{Cover1979} and the classic network
coding by Ahlswede, Cai, Li, and Yeung \cite{Yeung_networkcoding} to
noisy networks. The main idea of noisy network coding is to derive an
explicit expression of the achievable rate for each cut-set of the
network. Then, by comparing with the cut-set upper bound, noisy
network coding can be shown to achieve to within constant gap to the
capacity of general multisource multicast Gaussian networks.

A key assumption made in both \cite{Avestimehr_relaynetwork} and
\cite{Kim_noisy_network_coding} is that the noises in the Gaussian
relay network are independent with each other. This assumption may not
hold in practical systems, where common interferences from other
sources play a role. In this paper, we are interested in the following
question. In the context of Gaussian relay networks with correlated
noises, can noisy network coding achieve within constant bits to the
capacity as well? This paper gives a negative answer by studying a
four-node Gaussian chain network with correlated noises. It is shown
that, in a certain scenario, the noisy-network-coding rate (with
Gaussian input and Gaussian quantization) has an
unbounded gap to the cut-set bound, whereas a concatenation of a
modified {\it{correlation-aware}} noisy network coding strategy and a
conventional decode-and-forward scheme achieves to within half a bit of
the cut-set bound in this specific case.

\section{Channel Model}


\begin{figure} [t]
\centering
\includegraphics[width=3.3in]{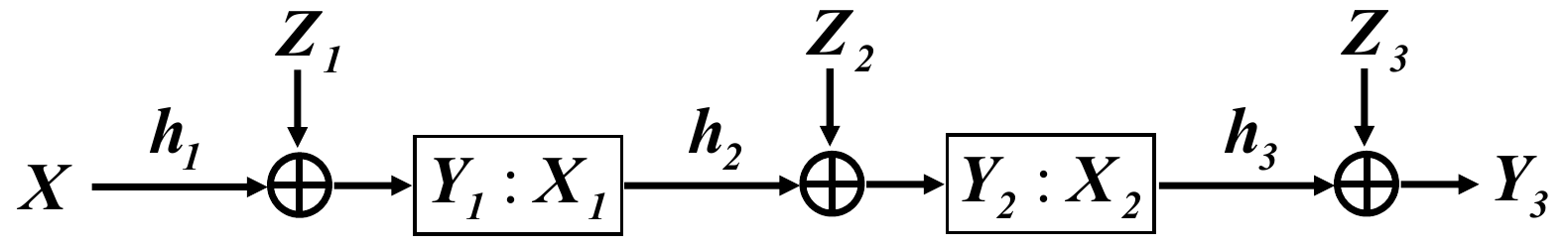}
\caption{A four-node Gaussian relay chain network} \label{line_relay}
\end{figure}

The four-node Gaussian relay chain, as depicted in
Fig.~\ref{line_relay}, consists of a source node, a destination node,
and two relay nodes. The source communicates with the destination with
the help from the two relays in between.
Information passes from the source to the neighboring relay, and to
the next, then finally to the destination. The input-output
relationship can be described as follows:  \begin{eqnarray}
Y_1 &=& h_1 X + Z_1, \nonumber \\
Y_2 &=& h_2 X_1 + Z_2, \nonumber \\
Y_3 &=& h_3 X_2 + Z_3. \nonumber
\end{eqnarray}
Without loss of generality, assume that the transmit power of all nodes are
normalized to one, and the variances of the receiver noises are also
normalized to one, i.e., $Z_i \sim \mathcal{N}(0, 1)$. The receiver
noises are i.i.d.\ in time, but the noises $[Z_1, Z_2, Z_3]$ are
correlated with the following correlation matrix:
\begin{equation}
K_Z = \left[  \begin{array}{ccc} 1 &  \rho_{12} &  \rho_{13}  \\
\rho_{12}  & 1 & \rho_{23}  \\
\rho_{13} &  \rho_{23} & 1
\end{array} \right], \nonumber
\end{equation}
where $K_z$ is positive semidefinite and $\rho_{ij}$ is the correlation
coefficient between $Z_i$ and $Z_j$. Note that the relay operation
must be causal in time.

\section{Suboptimality of the Noisy Network Coding}



We begin by showing that using noisy network coding with the choices of
Gaussian inputs and Gaussian quantization noises, the gap between the
achievable rate and the cut-set bound can be unbounded for a Gaussian relay
chain network with a certain noise correlation structure. First, an upper bound
to the cut-set bound of the four-node relay chain can be computed as follows:
\begin{eqnarray}
\lefteqn{\max_{p(x, x_1, x_2)} \min \{I(X; Y_1 Y_2 Y_3|X_1 X_2), I(XX_1; Y_2Y_3|X_2),} \nonumber \\
&&   \quad \quad \quad  I(XX_1X_2; Y_3), I(XX_2; Y_1 Y_3|X_1) \} \nonumber \\
&\le& \min \{ \max I(X; Y_1 Y_2 Y_3|X_1 X_2), \max I(XX_1; Y_2Y_3|X_2),\nonumber \\
&& \quad \quad \;  \max I(XX_1X_2; Y_3), \max I(XX_2; Y_1 Y_3|X_1)  \} \nonumber \\
&=& \min \{\overline{C}(\mathcal{S}_1), \overline{C}(\mathcal{S}_2),
\overline{C}(\mathcal{S}_3) ,\overline{C}(\mathcal{S}_4) \} \nonumber
\\
&\triangleq& \overline{C}\label{cut-set}
\end{eqnarray}
where the cut-sets are defined as $\mathcal{S}_1 =\{X\}$, $\mathcal{S}_2 =\{X, X_1\}$, $\mathcal{S}_3 =\{X, X_1, X_2\}$, and $\mathcal{S}_4 =\{X, X_2\}$,
and $\overline{C}(\mathcal{S}_i), i=1,2,3,4$ are the four cut-set upper bounds, which can be calculated as
\begin{eqnarray}
\overline{C}(\mathcal{S}_1) &=& \max I(X; Y_1 Y_2 Y_3|X_1 X_2) \nonumber \\
&=&\frac{1}{2} \log \left(1 + \frac{(1 - \rho_{23}^2)h_1^2}{|K_Z|} \right),
\end{eqnarray}
%
%
and
\begin{eqnarray}
\overline{C}(\mathcal{S}_2) &=& \max I(XX_1; Y_2Y_3|X_2) \nonumber \\
&=& \frac{1}{2} \log \left( 1 + \frac{h_2^2}{1- \rho_{23}^2} \right),
\label{cut-set-S2}
\end{eqnarray}
and
\begin{eqnarray}
\overline{C}(\mathcal{S}_3) &=& \max I(XX_1X_2; Y_3) \nonumber \\
&=& \frac{1}{2} \log(1 + h_3^2), \label{cut-set-S3}
\end{eqnarray}
and
\begin{eqnarray}
\overline{C}(S_4) &=& \max I(XX_2; Y_1Y_3|X_1) \nonumber \\
&=& \hf \log \left( 1 +\frac{h_1^2 + h_3^2 + h_1^2h_3^2}{1 - \rho_{13}^2}\right) \nonumber \\
&\ge& \overline{C}(\mathcal{S}_3),
\end{eqnarray}
which is redundant.

The main point of noisy network coding is that an
achievable rate can be derived for each of the cut-sets
$\mathcal{S}_1$, $\mathcal{S}_2$, and $\mathcal{S}_3$ using a
generalization of the compress-and-forward scheme. For convenience, we
state the achievable rates as follows.

\begin{theo}[Noisy Network Coding Theorem \cite{Kim_noisy_network_coding}]
Let $\mathcal{D}=\mathcal{D}_1=\mathcal{D}_2=\cdots=\mathcal{D}_N$. A rate
tuple $(R_1, \cdots, R_N)$ is achievable for the DMN $p(y^N|x^N)$ if there
exists some joint pmf $p(q)\prod_{k=1}^{N}p(x_k|q)p(\hat{y}_k|y_k, x_k, q)$
such that
\begin{eqnarray}
R(\mathcal{S}) &<& \min_{d \in \mathcal{\mathcal{S}}^c \cap \mathcal{D}}
I(X(\mathcal{S});
\hat{Y}(\mathcal{S}^c), Y_d|X(\mathcal{S}^c), Q) \nonumber \\
&&\quad - I(Y(\mathcal{S}); \hat{Y}(\mathcal{S})|X^N, \hat{Y}(\mathcal{S}^c),
Y_d, Q)
\end{eqnarray}
for all cutsets $\mathcal{S} \subseteq [1:N]$ with $S^c \cap \mathcal{D} \neq
\emptyset$, where $R(\mathcal{S}) = \sum_{k \in \mathcal{\mathcal{S}}}R_k$.
\end{theo}

Although the quantization in the above noisy network coding theorem can in
theory have arbitrary distributions, Gaussian inputs and Gaussian quantization
noises are usually adopted for Gaussian networks \cite{Wang_ReceiverCooperation}
\cite{zhou_z_relay}, and are shown to achieve constant gap to capacity for
networks with uncorrelated noises \cite{Kim_noisy_network_coding}. Thus, this
paper follows the same choice, i.e.,
\begin{eqnarray}
\hat{Y}_i = Y_i + \hat{Z}_i
\end{eqnarray}
where the quantization noise $\hat{Z}_i \sim \mathcal{N}(0,
\q_i)$ is independent with everything else. Now applying the
noisy network coding theorem with Gaussian inputs and Gaussian quantization
noises, the following achievable rates for cut-sets $\mathcal{S}_1$, $\mathcal{S}_2$,  and $\mathcal{S}_3$ can be derived:
\begin{eqnarray}
R(\mathcal{S}_1) &=& I(X; \hat{Y}_1 \hat{Y}_2 Y_3 |X_1 X_2) \nonumber \\
&=& h\left( \begin{array}{r}  h_1X + Z_1 + \hat{Z}_1  \\
Z_2 + \hat{Z}_2  \\
Z_3
\end{array} \right) - h\left( \begin{array}{r}  Z_1 + \hat{Z}_1  \\
Z_2 + \hat{Z}_2  \\
Z_3
\end{array} \right) \nonumber \\
&=& \hf \log \left(1+ \frac{(1 +  \q_2 -\rho_{23}^2
)h_1^2}{|K_\beta|} \right), \label{R1:noisynetworking}
\end{eqnarray}
where
\begin{equation}
|K_\beta| = \left | \begin{array}{ccc} 1 + \q_1 & \rho_{12} &
\rho_{13} \\ \rho_{12} & 1 +\q_2 & \rho_{23} \\ \rho_{13} &
\rho_{23} & 1  \end{array}\right|, \label{Kbeta}
\end{equation}
%
%
%
and
\begin{eqnarray}
R(\mathcal{S}_2) &=& I(XX_1; \hat{Y_2}Y_3|X_2) - I(Y_1; \hat{Y}_1|XX_1X_2 \hat{Y}_2 Y_3) \nonumber \\
&=& \frac{1}{2}\log\left(1+\frac{h_2^2}{1 + \q_2  - \rho_{23}^2}
\right) \nonumber \\
& & \qquad - \frac{1}{2}\log\left( 1+\frac{1 - \rho_{13}^2}{\q_1} \right).
\end{eqnarray}
and
\begin{eqnarray}
R(\mathcal{S}_3) &=& I(XX_1X_2; Y_3) - I(Y_1 Y_2;\hat{Y}_1 \hat{Y}_2|XX_1X_2Y_3) \nonumber \\
&=& \frac{1}{2} \log (1 + h_3^2) \nonumber \\
& & - \frac{1}{2}\log \frac{\left|  \begin{array}{cc} 1- \rho_{13}^2+ \q_1 &  \rho_{12}-\rho_{13}\rho_{23} \\
\rho_{12}-\rho_{13}\rho_{23} & 1- \rho_{23}^2 + \q_2
\end{array} \right|}{\q_1 \q_2}.
\label{R3:noisynetworking}
\end{eqnarray}

The achievable rate is then upper bounded by the minimum of the three:
\begin{equation}
R \le \min\{R(\mathcal{S}_1), R(\mathcal{S}_2),R(\mathcal{S}_3) \}.
\label{R:noisynetworkcoding}
\end{equation}

Now, consider a special scenario when the noise $Z_3$ is independent with both
$Z_1$ and $Z_2$, and channel strengths $h_2^2$ and $h_3^2$ scale with $h_1^2$,
i.e.,
\begin{equation} \label{4node_scenario_1}
\rho_{13} = \rho_{23} = 0,
\end{equation}
and
\begin{equation} \label{4node_scenario_2}
h_2^2 =h_3^2 =  \frac{h_1^2}{1 - \rho_{12}^2}.
\end{equation}
This special setting gives us the following cut-set bounds:
\begin{equation} \label{cut-set_4node}
\overline{C}(\mathcal{S}_1) = \overline{C}(\mathcal{S}_2)
=\overline{C}(\mathcal{S}_3) = \hf \log \left( 1+ \frac{h_1^2}{1 - \rho_{12}^2}
\right),
\end{equation}
and the following achievable rates for cut-sets $\mathcal{S}_1$ to
$\mathcal{S}_3$:
\begin{eqnarray}
R(\mathcal{S}_1) &=& \hf \log \left(1 +\frac{h_1^2}{1 + \q_1-\frac{\rho_{12}^2}{1 + \q_2}} \right), \nonumber \\
R(\mathcal{S}_2) &=& \hf \log \left(1+ \frac{h_1^2}{(1 - \rho_{12}^2)(1 +
\q_2)} \right) \nonumber \\
&&- \hf \log \left(1+\frac{1}{\q_1} \right), \nonumber \\
R(\mathcal{S}_3) &=& \hf \log\left(1+ \frac{h_1^2}{1 - \rho_{12}^2} \right) \nonumber \\
&&- \hf \log \left(1 + \frac{\q_1 + \q_2 + 1 -
\rho_{12}^2}{\q_1\q_2} \right). \nonumber
\end{eqnarray}

Next, we show that, the gaps $\overline{C}(\mathcal{S}_1) -
R(\mathcal{S}_1)$, $\overline{C}(\mathcal{S}_2) - R(\mathcal{S}_2)$,
$\overline{C}(\mathcal{S}_3) - R(\mathcal{S}_3)$ {\it{cannot}} be made all
finite when $\rho_{12}^2 \rightarrow 1$. First, the gap on the cut-set
$\mathcal{S}_1$ is given by
\begin{eqnarray}
\Delta(\mathcal{S}_1)&=& \overline{C}(\mathcal{S}_1) - R(\mathcal{S}_1) \nonumber \\
&=& \hf \log \left(1+ \frac{h_1^2}{1 - \rho_{12}^2} \right) \nonumber \\
&& - \hf \log \left(1+\frac{h_1^2}{1 + \q_1-\frac{\rho_{12}^2}{1 + \q_2}} \right), \label{Delta_S1}
\end{eqnarray}
and the gap on the cut-set $\mathcal{S}_2$ is lower bounded by
\begin{eqnarray}
\Delta(\mathcal{S}_2) &=& \overline{C}(\mathcal{S}_2) - R(\mathcal{S}_2) \nonumber \\
&=& \hf \log \left(1+ \frac{h_1^2}{1- \rho_{12}^2} \right) + \hf \log \left(1+\frac{1}{\q_1} \right)  \nonumber \\
&&- \hf \log\left(1+ \frac{h_1^2}{(1 - \rho_{12}^2)(1 + \q_2)} \right) \nonumber \\
&\ge& \hf \log \left(1+\frac{1}{\q_1} \right),
\end{eqnarray}
and the gap on the cut-set $\mathcal{S}_3$ is lower bounded by the same number
as well, i.e.,
\begin{eqnarray}
\Delta(\mathcal{S}_3)&=& \overline{C}(\mathcal{S}_3) - R(\mathcal{S}_3) \nonumber \\
&=& \hf \log \left(1 + \frac{\q_1 + \q_2 + 1 - \rho_{12}^2}{\q_1\q_2} \right) \nonumber \\
&\ge& \hf \log \left(1 + \frac{1}{\q_1}\right).
\end{eqnarray}

Now, since
\begin{equation}
\overline{C} - R  \ge \max \{\Delta(\mathcal{S}_1), \Delta(\mathcal{S}_2), \Delta(\mathcal{S}_3) \},
\end{equation}
in order to make $\overline{C} - R $ finite, all three gaps have to be upper bounded by a finite number. 
Inspecting the gap of $\Delta(\mathcal{S}_1)$ in (\ref{Delta_S1}), in order to make it finite when $\rho_{12} \rightarrow \infty$, both $\mathsf{q}_1$ and $\mathsf{q}_2$ have to go to zero.  However, in this case, $\Delta(\mathcal{S}_2)$ and $\Delta(\mathcal{S}_3)$ are apparently unbounded. Therefore, in the
scenario of (\ref{4node_scenario_1}) and (\ref{4node_scenario_2}), as
$\rho_{12}^2$ goes to $1$, it is impossible to keep all three gaps
$\Delta(\mathcal{S}_1)$, $\Delta(\mathcal{S}_2)$, and $\Delta(\mathcal{S}_3)$
finite simultaneously. As a consequence, for the four-node Gaussian
chain network with correlated noises, the noisy network coding achievable rate with the choice of Gaussian inputs and Gaussian quantization noises has an unbounded gap to the cut-set upper bound.

\section{An Optimal Concatenated Scheme}
It is known that the cut-set upper bound is not always tight for the relay
channel \cite{marko, Zhang_partial_converse}, but for the four-node Gaussian
chain network, does the cut-set bound have an infinite gap to capacity? Or, is
it the noisy networking coding achievable rate that has an infinite gap to
capacity?
To answer this question, we show in the following that the cut-set bound
(\ref{cut-set_4node}) can actually be achieved to within half a bit for this
four-node relay network with the particular noise correlation structure
(\ref{4node_scenario_1}) by a simple concatenation of a correlation-aware noisy network coding strategy and a conventional
decode-and-forward scheme. This justifies the suboptimality of the noisy
network coding for Gaussian relay networks with correlated noises.

Inspecting the structure of the four-node relay network in
Fig.~\ref{line_relay} and the special correlation structure
(\ref{4node_scenario_1}), 
it is easy to see that the last node is essentially
independent of the first three nodes in this example.  Now the first
three nodes (from the source node $X$ to the second relay node $Y_2$)
forms a three-node Gaussian relay channel, so we can apply the noisy
network coding theorem just to the first three nodes.  With the source
message decoded at $Y_2$, the second relay node can then re-encode the
source information and forward to the destination $Y_3$.
With $Y_1$ serving as a noisy-network-coding type of relay and $Y_2$
serving as a decode-and-forward type of relay, this is essentially a
concatenation of the noisy network coding and the decode-and-forward
scheme. The achievable rate of this concatenated scheme can be derived
as follows.

In the first relaying stage where $Y_1$ serves as a
noisy-network-coding type of relay, according to the noisy network
coding theorem \cite[Theorem~1]{Kim_noisy_network_coding}, $Y_2$ can
decode the source message if the following rate is satisfied:
\begin{eqnarray}
R &\le&\min \{I(X,X_1; Y_2)-I(Y_1;\hat{Y}_1|X, X_1, Y_2) , \nonumber \\
&& \quad \quad \; I(X; Y_2, \hat{Y}_1|X_1)\},
\end{eqnarray}
for some distribution
\begin{equation}
p(x, x_1, y_1, \hat{y}_1) = p(x)p(x_1)p(y_1|x, x_1)p(\hat{y}_1|x_1, y_1). \nonumber
\end{equation}
Substituting Gaussian inputs $X \sim \mathcal{N}(0, 1)$, $X_1 \sim \mathcal{N}(0, 1)$, and Gaussian quantization signal $\hat{Y}_1 = Y_1 + \hat{Z}_1$, where $\hat{Z}_1 \sim \mathcal{N}(0, \q_1)$ is independent with everything else, we have the following achievable rate in the first stage:
\begin{eqnarray} \label{CF_DF_rate_1}
R &\le& \min \left\{ \hf \log \left(1 + \frac{h_1^2}{1 - \rho_{12}^2}  \right)
- \hf \log \left( 1 + \frac{\q_1}{1- \rho_{12}^2} \right), \right.\nonumber \\
&& \left. \quad \quad \;\; \hf \log (1 + h_2^2) - \hf \log \left(1 + \frac{1-\rho_{12}^2}{\q_1} \right) \right\}.
\end{eqnarray}

Next, with the source message decoded at $Y_2$, the second relay node
acts as a decode-and-forward type of relay, which re-encodes and
forwards the source message to the destination $Y_3$ through
the Gaussian channel of channel gain $h_3$. The destination can
successfully decode the source message if
\begin{eqnarray}
R \le  \hf \log(1+ h_3^2). \label{CF_DF_rate_2}
\end{eqnarray}

Combining the above rate constraints (\ref{CF_DF_rate_1}) and
(\ref{CF_DF_rate_2}) gives us the following achievable rate by the
concatenated scheme:
\begin{eqnarray} \label{CF_DF_rate}
R &\le& \min \left\{ \hf \log \left(1 + \frac{h_1^2}{1 - \rho_{12}^2}  \right)
- \hf \log \left( 1 + \frac{\q_1}{1- \rho_{12}^2} \right), \right.\nonumber \\
&&  \quad \quad \;\; \hf \log (1 + h_2^2) - \hf \log \left(1 + \frac{1-\rho_{12}^2}{\q_1} \right), \nonumber \\
&& \quad \quad \;\; \left.  \hf \log(1+ h_3^2) \right\}.
\end{eqnarray}
Comparing the above achievable rate with the cut-set upper bound with $\rho_{12}$ and $\rho_{23}$ set to zero:
\begin{eqnarray}
\overline{C} &=& \min \left \{\hf \log \left(1+\frac{h_1^2}{1 - \rho_{12}^2} \right),
\hf \log(1 +h_2^2) , \right. \nonumber \\
&& \quad\quad\;\; \left. \hf \log(1+ h_3^2) \right\},
\end{eqnarray}
we have the difference upper bounded by
\begin{equation} \label{Concatenation_Gap}
\overline{C} - R \le \max \left\{ \hf \log \left( 1 + \frac{\q_1}{1- \rho_{12}^2} \right), \hf \log \left(1 + \frac{1-\rho_{12}^2}{\q_1} \right) \right\}
\end{equation}
It is easy to see that the first term monotonically increases with $\q_1$ while the second term monotonically decreases. As a result, to minimize the maximum of the two terms, we need
\begin{eqnarray}
\hf \log \left(1 + \frac{1-\rho_{12}^2}{\q_1^*} \right) = \hf \log \left( 1+ \frac{\q_1^*}{1- \rho_{12}^2} \right),
\end{eqnarray}
which results in the optimal {\it{correlation-aware}} quantization
level $\q_1^* = 1- \rho_{12}^2$. Substituting $\q_1^*$ into
(\ref{Concatenation_Gap}) gives us $\overline{C} - R < \hf$.
Therefore, for the four-node Gaussian chain
network as shown in Fig.~\ref{line_relay}, in the scenario where $\rho_{13}=\rho_{23}=0$, the cut-set upper bound
can be achieved to within constant gap. 

\begin{figure} [t]
\centering
\includegraphics[width=3.4in]{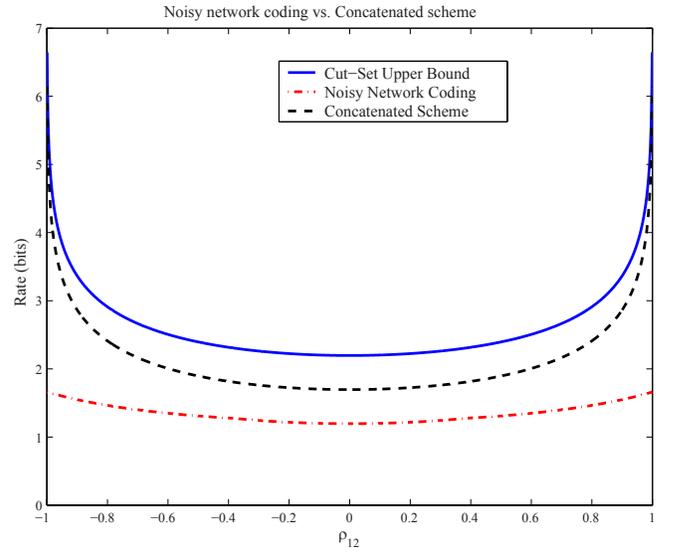}
\caption{Noisy network coding vs. concatenated scheme}
\label{Fig_NNC}
\end{figure}

Fig.~\ref{Fig_NNC} shows a numerical example for comparing noisy
network coding and the concatenated scheme. In the simulation, we let
$h_{1}^2 = 20$dB and all other channel parameters are set to satisfy
(\ref{4node_scenario_1}) and (\ref{4node_scenario_2}). For the
quantization parameters, we choose the optimal quantization level
$\q_1 = 1 - \rho_{12}^2$ and let $\q_2 = 1$. As can be seen from the figure, when $\rho_{12}$ approaches to $+1$ or $-1$, both the cut-set upper bound and the achievable rate by the concatenated scheme go to infinity. However, the achievable rate by the noisy network coding scheme remains finite, making the gap to the cut-set bound unbounded.

\section{Conclusion}
This paper studies the optimality of the noisy network coding for a
four-node Gaussian chain network with correlated noises. It is shown
that, under a certain noise correlation structure, noisy network
coding with Gaussian inputs and Gaussian quantization
noises has an infinite gap to the cut-set upper bound. But, the upper
bound can be achieved to within half a bit in this specific case by a simple concatenation of a correlation-aware noisy network coding strategy and a decode-and-forward scheme.

\bibliographystyle{IEEEtran}
\bibliography{IEEEabrv,../ref/main}

\begin{thebibliography}{1}
\providecommand{\url}[1]{#1}
\csname url@samestyle\endcsname
\providecommand{\newblock}{\relax}
\providecommand{\bibinfo}[2]{#2}
\providecommand{\BIBentrySTDinterwordspacing}{\spaceskip=0pt\relax}
\providecommand{\BIBentryALTinterwordstretchfactor}{4}
\providecommand{\BIBentryALTinterwordspacing}{\spaceskip=\fontdimen2\font plus
\BIBentryALTinterwordstretchfactor\fontdimen3\font minus
  \fontdimen4\font\relax}
\providecommand{\BIBforeignlanguage}[2]{{%
\expandafter\ifx\csname l@#1\endcsname\relax
\typeout{** WARNING: IEEEtran.bst: No hyphenation pattern has been}%
\typeout{** loaded for the language `#1'. Using the pattern for}%
\typeout{** the default language instead.}%
\else
\language=\csname l@#1\endcsname
\fi
#2}}
\providecommand{\BIBdecl}{\relax}
\BIBdecl

\bibitem{Avestimehr_relaynetwork}
S.~Avestimehr, S.~Diggavi, and D.~Tse, ``Wireless network information flow: a
  deterministic approach,'' \emph{Submitted to IEEE Trans. Inf. Theory}, 2009.

\bibitem{Chang_constantbitrelay}
\BIBentryALTinterwordspacing
W.~Chang, S.-Y. Chung, and Y.~H. Lee, ``Gaussian relay channel capacity to
  within a fixed number of bits,'' 2010. [Online]. Available:
  \url{http://arxiv.org/abs/1011.5065}
\BIBentrySTDinterwordspacing

\bibitem{Kim_noisy_network_coding}
S.-Y. Lim, Y.-H. Kim, A.~El~Gamal, and S.-Y. Chung, ``Noisy network coding,''
  \emph{Submitted to IEEE Trans. Inf. Theory}, 2010.

\bibitem{Cover1979}
T.~M. Cover and A.~El\hspace{1mm}Gamal, ``{Capacity theorems for the relay
  channel},'' \emph{{IEEE} Trans. Inf. Theory}, vol.~25, no.~5, pp. 572--584,
  Sep. 1979.

\bibitem{Yeung_networkcoding}
R.~Ahlswede, N.~Cai, R.~Li, and R.~W. Yeung, ``{Network information flow},''
  \emph{{IEEE} Trans. Inf. Theory}, vol.~46, pp. 1004 --1016, Jul 2000.

\bibitem{zhou_z_relay}
\BIBentryALTinterwordspacing
L.~Zhou and W.~Yu, ``{Gaussian z-interference channel with a relay link:
  achievability region and asymptotic sum capacity},'' 2010. [Online].
  Available: \url{http://arxiv.org/abs/1006.5087}
\BIBentrySTDinterwordspacing

\bibitem{marko}
M.~Aleksic, P.~Razaghi, and W.~Yu, ``{Capacity of a class of modulo-sum relay
  channels},'' \emph{{IEEE} Trans. Inf. Theory}, vol.~55, no.~3, pp. 921--930,
  Mar. 2009.

\bibitem{Zhang_partial_converse}
Z.~Zhang, ``{Partial converse for a relay channel},'' \emph{{IEEE} Trans. Inf.
  Theory}, vol.~34, no.~5, pp. 1106--1110, Sep. 1988.

\end{thebibliography}

\end{document}